# A Dual-Purpose Ion-Accelerator for Nuclear-Reaction-Based Explosives-and SNM-Detection in Massive Cargo

*M.B. Goldberg [a] [\*], V. Dangendorf [b], D. Vartsky [a], D. Bar [a], R. Böttger [b], M. Brandis [a],*

*B. Bromberger [b], G. Feldman [a], E. Friedman [c], D. Heflinger [a], R. Lauck [b], S. Löb [b],*

*P. Maier-Komor [d], I. Mardor [a], I. Mor [a], K.-H. Speidel [e], K.Tittelmeier [b], M. Weierganz [b]*

[a] Soreq Nuclear Research Center, Route 4111 West, 81800 Yavne, Israel

[b] Physikalisch-Technische Bundesanstalt (PTB), Bundesallee 100, 38116 Braunschweig, Germany

[c] Racah Institute of Physics, the Hebrew University of Jerusalem, 91904 Jerusalem, Israel

[d] Physik-Department E12, Technische Universität München, 85747 Garching, Germany

[e] Helmholtz-Institut für Strahlen- und Kernphysik, Nussallee 14-16, 53115 Bonn, Germany

## Abstract

A concept is presented for a dual-purpose ion-accelerator, capable of serving as radiation source in a versatile, nuclear-reaction-based inspection system for massive-cargo. The system will automatically and reliably detect small, operationally-relevant quantities of concealed explosives and Special Nuclear Materials (SNM). It will be cost-effective, employing largely-common hardware, but different reactions and data acquisition modes. Typical throughput is expected to be 10-20 aviation containers/hr, at the beam intensities specified below.

With such an inspection system, explosives are detected via $\gamma$-Resonance Absorption (GRA) in $^{14}N$ using 9.17 MeV $\gamma$-rays produced in $^{13}C(p,\gamma)$, and SNM via Dual-Discrete-Energy $\gamma$-Radiography (DEGR) with 15.11 & 4.43 MeV $^{12}C$ $\gamma$-rays from $^{11}B(d,n)$. Simultaneously with the scan, 1-17 MeV neutrons from the latter reaction will yield complementary information, both on explosives and on SNM, via Fast-Neutron Resonance Radiography (FNRR). Few-view radiography will be implemented throughout, since spatial reconstruction of threat-object densities reduces false-alarm rates drastically.

Nevertheless, if a cargo item does alarm the system on SNM, confirmation of its presence and composition will be effected via a secondary-screening technique, namely, induced-fission decay-signatures, employing the $^{11}B(d,n)$ neutrons. This should only be required in solitary cases and will thus not impede cargo flow to any appreciable extent. For explosives, the GRA/FNRR combination comprehensively covers the entire spectrum of substances in the arena and no secondary-screening technique should be required.

The essence of the accelerator concept is a fixed-energy machine, alternately delivering mass-2 beams of $H_2^+$ (3 mA, cw) and deuterons (0.2 mA, pulsed) for GRA and DEGR/FNRR, respectively. It will operate at precisely double the GRA resonance energy of $E_p$=1.746 MeV (namely, 3.492 MeV) and require beam-energy resolution no better than ~15 keV (FWTM).

This specification was confirmed in a recent measurement, first reported here, of the GRA emission-linewidth obtained with $H_2^+$ ions, when driving the resonance into the depth of a moderately-thick $^{13}C$ target. For most acceleration techniques, such beam-energy resolution requirements are not unduly stringent, which works in favour of the high-current requirement. On deuteron beams there are no energy resolution constraints, as the $^{11}B(d,n)$ reaction is non-resonant.

[\*] on sabbatical leave at PTB-Braunschweig

## 1. Introduction

The threat to civil aviation, international trade and homeland security posed by illicitly-transported explosives and fissile materials is of ever-increasing concern. Moreover, contemporary terror organizations have become highly skilled in devising bombs that are smaller, more potent than hitherto and, no less significantly, progressively harder to detect.

In response, the U.S. Government passed a Bill in 2007 [1], which mandates that, within five years, 100% of all U.S.-bound maritime cargo, as well as all aviation cargo loaded onto passenger aircraft, be scanned for the above threat materials in foreign ports prior to shipment. Indeed, this implies that a very large number of inspection systems with capabilities beyond the present state-of-the-art will be needed, at a global hardware outlay estimated [ibid] to exceed 5000M$ (more than 1000 systems, at ~5M$ each). Clearly, a large fraction of the cost to acquire and operate these systems will fall on countries exporting to the U.S. Moreover, if the U.S. becomes less vulnerable, groups intent on harm will seek "softer" targets elsewhere.

Hence, inspection systems capable of rapid, reliable and automatic detection of operationally-relevant amounts of explosives and SNM concealed in massive cargo items are called for. A combined cargo screener concept that responds to this need is presented here. It is based on proven technologies and satisfies the criteria of: **1)** cost-effectiveness; **2)** high sensitivity and specificity to the full range of threat materials in the contemporary arena; **3)** low false-alarm rate; **4)** effective further screening of items that are not cleared on the first pass (system is truly "multi-level" [2]); **5)** low vulnerability to countermeasures aimed at defeating detection.

Within the R&D community, the realization is gradually taking hold that nuclear-reaction-based γ-ray and fast-neutron probes at judicially-chosen energies [3-17] are probably unique in providing the requisite sensitivities and specificities to both SNM and all types of explosives employed nowadays. In this context, a principal problem still to be overcome pertains to the dedicated, versatile MeV-range ion-accelerator that, together with a target, will constitute the radiation source. Indeed, such machines, hitherto developed almost exclusively for basic research facilities, have not yet attained adequate user-friendliness, reliability, robustness and cost-effectiveness for field applications under harsh environmental conditions.

Moreover, in contrast to X-ray-based systems, accumulating the required counting statistics within acceptable inspection times is contingent on the capability of the accelerator to deliver beams in the mA-intensity range. Even for low-energy ions, this technological problem has not yet been satisfactorily solved for field applications. Thus, the challenge at present is to conceive a system that reduces, as much as possible, the demands on such an accelerator, which will represent a large fraction of the cost of an operational system anyway.

## 2. The Concept

### 2.1 Dedicated Accelerator

The essence of the concept is a fixed-energy accelerator for mass-2 ions, alternately delivering beams of 3 mA (cw) $H_2^+$ to a $^{13}C$ target and 0.2 mA (pulsed) deuterons to a $^{11}B$ target, for GRA and DEGR/FNRR, respectively. It will operate at precisely double the GRA resonance energy of $E_p$=1.746 MeV (namely, 3.492 MeV). Minimal energy variability (±20 keV) will be required to optimize GRA resonance yields and compensate target-wear effects. For DEGR/FNRR, this is of no consequence, since the $^{11}B$(d,n) reaction is non-resonant.

The viability of this dual-purpose concept is non-trivial, since an appreciable contribution of proton-proton repulsion in $H_2^+$ ions to emission-line broadening of resonant 9.17 MeV $\gamma$-rays would prove seriously detrimental to GRA-system performance (see sections 4 & 5 below). This contribution, recently measured at the PTB Van-de-Graaff and first reported here, was found to be small compared to that observed when the level is populated by protons [18-20].

## 2.2 Combined Screening System

In the inspection process, bulk and sheet explosives are detected via $\gamma$-Resonance Absorption (GRA) in $^{14}$N [20] with 9.17 MeV $\gamma$-rays produced in $^{13}$C(p,$\gamma$), whereas SNM is detected via Dual-Discrete-Energy $\gamma$-Radiography (DEGR) with 15.11 & 4.43 MeV $^{12}$C $\gamma$-rays from $^{11}$B(d,n) [10,11]. Simultaneously with DEGR, 1-17 MeV neutrons from the latter reaction will yield complementary information, both on explosives and on SNM, via Fast-Neutron Resonance Radiography (FNRR) [3-7]. Few-view radiography [21] will be implemented throughout, since spatial reconstruction of threat-object densities reduces false-alarm rates drastically. The techniques and their underlying physical principles are outlined in section 3 below.

For cargo items that alarmed in the primary inspection mode, confirmation of SNM presence and its composition will be effected via Induced-Fission Decay-Signatures [13,14]. This will only be required in singular cases and will thus not impede cargo flow to any appreciable extent. For explosives detection, GRA and FNRR complement each other well and their combination comprehensively covers the entire spectrum of substances and cargo types in the arena (from palletized aviation cargo up to full-size marine containers). Alarms that are nevertheless registered will, in all likelihood, be resolved by taking one or two extra radiographic projections or by repeating the scan with higher counting statistics. Thus, no independent, secondary inspection technique should be required. A block diagram of the accelerator and combined cargo inspection system is shown below in Fig. 1.

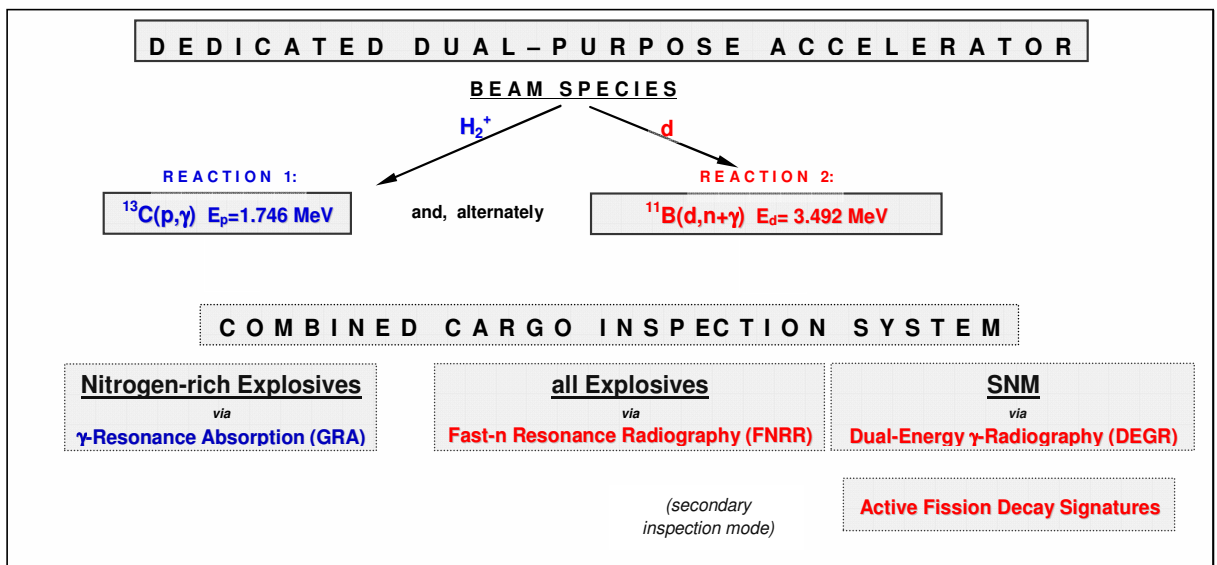

*Fig. 1: Block diagram of accelerator and combined cargo inspection system*

## 3. Screening Techniques

### 3.1 Dual-Discrete-Energy γ-Radiography (DEGR)

To penetrate massive cargo containers and detect the threat objects of interest, recourse must be made to characteristic attenuation features that come into play at photon energies in the 1-30 MeV range. At such energies, pair-production comes into play and the mass attenuation coefficient exhibits a rise above ~4 MeV, that is progressively more pronounced as the Z of the absorber increases. Thus, high-Z materials can be detected and distinguished, as a category, from low-Z and medium-Z substances that constitute the majority of benign transported items (Fig. 2). In practice, one (essentially normalizing) measurement needs to be performed at the global absorption minimum for all atomic numbers Z (between 1 and 4 MeV) and another at the highest energy possible [10,11], nuclear reaction yields permitting.

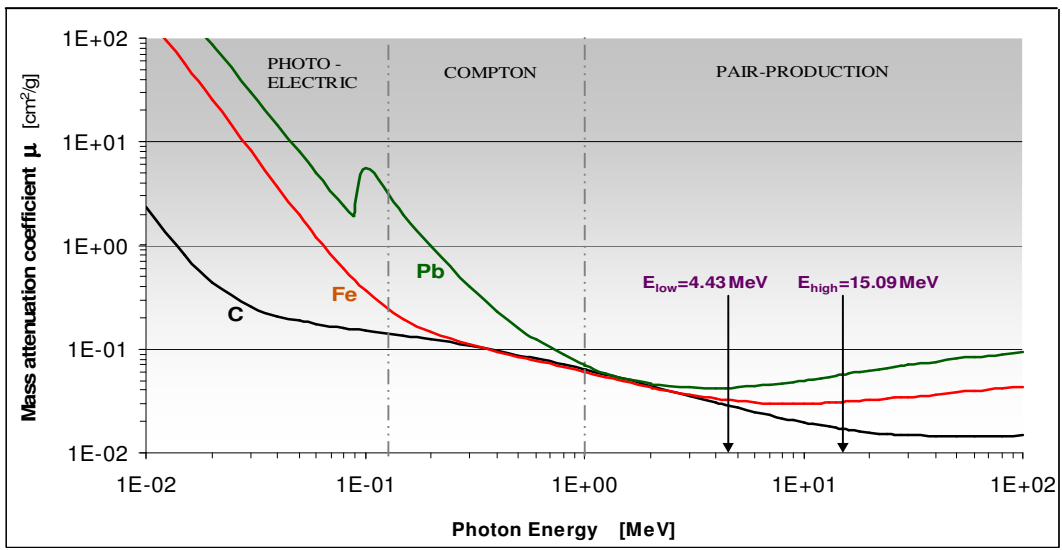

*Fig. 2: Mass attenuation coefficient as f (E), for low-, medium- and high-Z materials. The relevant $^{12}C$ γ-ray energies produced in $^{11}B(d,n)$ are marked with arrows.*

Moreover, SNM can be distinguished from benign, high-Z materials (Hg, Tl, Pb, Bi) [10] if at least 2 DEGR views are taken and 3-D densities reconstructed, even crudely (see Fig. 3).

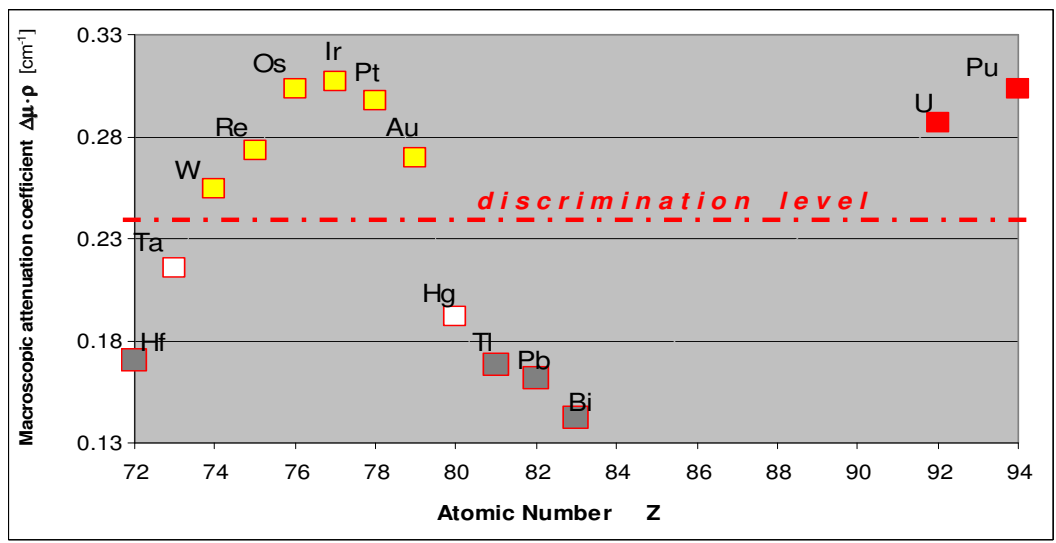

*Fig. 3: Macroscopic attenuation coefficient as f (Z) for heavy elements, calculated at $E_\gamma^{low}$=4.43 MeV and $E_\gamma^{high}$=15.11 MeV.*

Clearly, SNM (full red squares) can be distinguished from Rare-Earths, heavy stable elements (grey-centred squares) and, to a lesser extent, from transition metals Ta, Hg (white-centred squares). The main difficulty seems to be with noble metals, their neighbours (yellow-centred squares) and uranium (any isotopic composition), but these are unlikely to figure as undeclared cargo. Indeed, there is a definite interest in their being detected if illicitly-transported, so that the presence of such materials will not constitute a false-alarm, in the operational sense.

To construct a practicable system with good performance characteristics, judicious choices of γ-ray energies and populating reactions employing low-energy beams at sub-mA intensities are essential. To this end, a comprehensive literature study of reaction-induced, thick-target γ-yields has been performed [22,23]. Its conclusions on the most promising radiation source for the DEGR application are clear-cut [10]. They pertain to the $^{12}$C γ-lines, $E_\gamma^{low}$=4.43 MeV and $E_\gamma^{high}$=15.11 MeV, populated, in decreasing yield order, by the $^{11}$B(d,n), $^{13}$C($^3$He,α) and $^{10}$B($^3$He,p) reactions at energies $E_{beam}$<6 MeV. Moreover, the $^{11}$B(d,n) γ-spectra are also by far the cleanest, after fast-neutron events have been rejected (by shielding, by time-of-flight – TOF, or by pulse-shape-discrimination – PSD), so this is obviously the reaction of choice.

For $E_d$=3.5 MeV on thick $^{11}$B targets, γ-yields/deuteron are: ~3·10$^{-5}$ ($E_\gamma^{low}$) and ~7·10$^{-6}$ ($E_\gamma^{high}$)

The key issue is whether such a nuclear-reaction-based radiation source competes favourably with Bremsstrahlung (B-S) sources. Indeed, a semi-quantitative comparison was made to 5 & 9 MeV B-S [10]. Its salient features (for 200 μA, 3.5 MeV deuteron beams) are as follows:

1. Unfiltered B-S yields are a factor of ~100 higher
2. B-S spectra must be heavily filtered, to remove low-energy X-rays that reduce contrast
3. Minimally-filtered B-S yields are only a factor of ~5 higher
4. $^{11}$B(d,n+γ) Contrast-Sensitivity (=Δμ/μ) is a factor of ~ 5 higher
5. Overall figure-of-merit (FOM) = Contrast-Sensitivity · $\sqrt{\text{Yield}}$
6. On overall FOM, $^{11}$B(d,n+γ) has the edge over 5 & 9 MeV B-S

A preliminary measurement at PTB, using organic liquid scintillators in event-counting mode, has shown that, for calibrated graphite (C), iron (Fe) and lead (Pb) absorbers, the theoretical contrast sensitivities are experimentally well reproduced, if sufficiently narrow integration limits are set on the respective Compton edges in the γ-ray pulse-height spectrum.

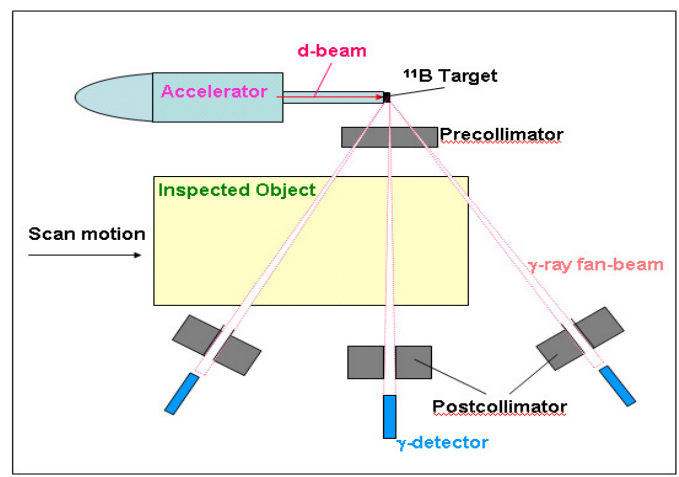

*Fig. 4: Schematic showing how few-view DEGR (and FNRR) Radiography can be performed in one pass, with multiple detector arrays.*

Simulations show[10] that a few-view DEGR system (Fig. 4) permits reliable detection of small, operationally-relevant SNM quantities even in massive cargo items that exhibit high image clutter. For deuteron beams of 200 µA, cargo throughout is estimated at ~20 aviation containers/hr. Radiation doses to inspected items are around 1-10 µGy. If the fast neutrons are largely shielded out, doses will decrease by approximately an order of magnitude.

### 3.2 Fast-Neutron Resonance Radiography (FNRR)

Fast-Neutron Resonance Radiography (FNRR) [5,7] is an imaging method that exploits the characteristic cross-section structure (resonances) of different isotopes in the energy-range $E_n$=1-10 MeV. The pulsed FNRR method, originally named PFNTS, was first proposed and studied by the Oregon University group [3] and later by Tensor-Technology, Inc. [4,24]. The method holds promise for identifying and detecting a broad range of explosives due to its ability to determine simultaneously the identity and density distribution of the principal elements present in explosives, such as C, O and N.

In PFNTS the inspected object is irradiated with pulsed neutrons of a broad spectral distribution in the above energy range. If the inspected object contains elements that exhibit sharp cross-section resonances, the neutron transmission spectrum will be modified such that it will exhibit dips and peaks at specific energies corresponding to these resonances. The spectrum thus carries information on the elemental composition of the object. Fig. 5 shows a calculated transmission through 10 cm thick objects comprising TATP (an improvised explosive), water and polyethylene, all of which have similar densities.

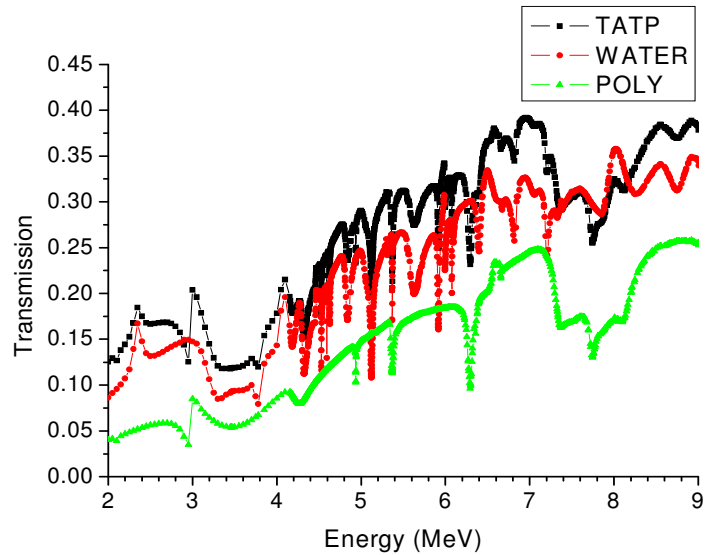

*Fig. 5: Calculated transmission through 10 cm thick TATP, water and polyethylene*

Indeed, these items would look the same to an X-ray probe, which measures density only, whereas FNRR determines the characteristic stoichiometric ratio of elements in the substance. Thus, the neutron transmission spectrum can be used to distinguish TATP and other explosives from benign materials of similar physical density. A pre-requisite for FNRR is the precise knowledge of the neutron energy. In PFNTS this is achieved by Time-of-Flight (TOF) spectroscopy, requiring pulsed beams and imaging systems with TOF measurement capability.

Unlike conventional X-ray radiography systems, FNRR does not rely on the skills of a human operator to identify the threat object. Instead, it relies on automatic identification of the concealed contraband, via few-view spatial reconstruction of its elemental composition. FNRR can detect reliably standard and improvised explosives in solid or liquid form.

The early PFNTS detectors developed by Oregon University and Tensor Inc. had pixel sizes of several centimeters. This posed an intrinsic limitation on the position resolution, which did not permit reliable detection of thin sheet explosives. Thus, since 2004, PTB, Soreq NRC and the Weizmann Institute of Science have been developing a sub-mm spatial resolution FNRR system based on time-resolved, integrative, optical neutron detectors (TRION) [7, 25-30].

Fig. 6 shows the latest TRION version, which was operated recently in a demonstration experiment. It is capable of acquiring simultaneously up to 4 neutron transmission images at different, predefined energy windows. A detailed description of its operating principle and performance can be found in a recent publication [31]. Fig. 7 illustrates the typical spatial resolution (~0.5mm) obtainable with this detector. The image shows a fast neutron radiograph of a gun, its magazine, a vial of UO₂ powder and a bar of tungsten (W). Image quality is suitable for visual inspection, in addition to the automatic detection based on elemental information.

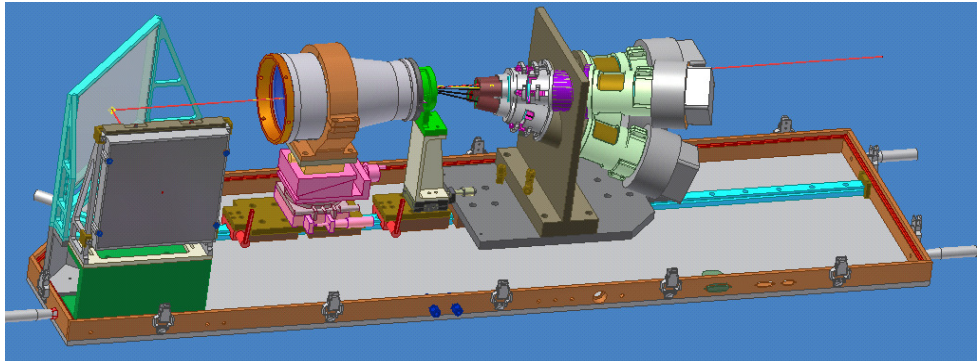

*Fig. 6:   Engineering drawing of TRION with 4 independently-gateable cameras for simultaneous imaging at 4 different neutron energies*

As mentioned above, an FNRR transmission spectrum carries information about the elemental composition of the object. Fig. 8a shows a picture of 6 milk bottles and 4 small vials. The content of the upper-middle bottle was replaced by a simulant of TATP, an improvised explosive frequently used by terrorists. The vials contain various mixtures of TATP-simulant and water. Fig. 8b shows a conventional γ-ray and neutron radiograph obtained using the TRION detector. As can be observed, the radiograph does not reveal any difference between milk and TATP.

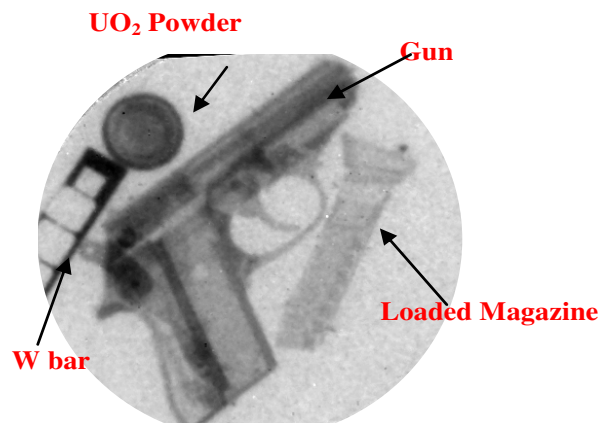

*Fig. 7:  TRION fast neutron radiograph*

However, taking into account the information provided by FNRR, one can determine the composition of the elements reconstruct their areal densities. Fig. 8c shows the reconstructed FNRR image. Here, only the bottle and vials containing TATP are visible. On the basis of this information, the system can alarm automatically and point to the suspect region.

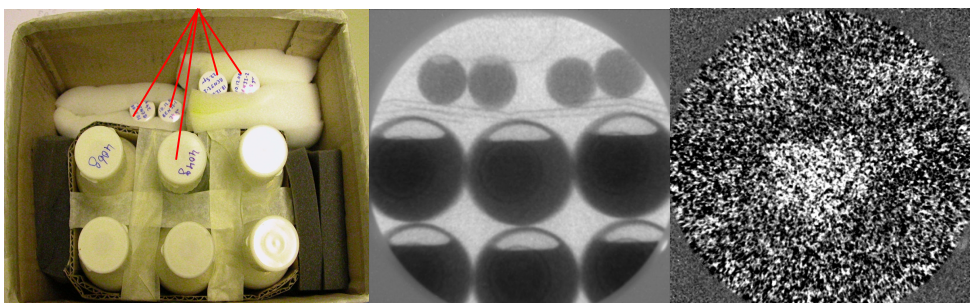

*Fig. 8:   a) Photograph of milk and TATP bottles   b) combined γ-neutron radiographic image;   c) reconstructed FNRR image.*

Compared to early PFNTS detectors, TRION detectors combine excellent spatial resolution with adequate timing (energy) resolution. Time-resolved optical readout permits construction of large-area detectors with relatively high (15-20%) detection efficiency, as well as good discrimination between γ-rays and neutrons by TOF spectrometry.

Whereas the 2nd generation TRION detector described above can capture up to 4 images at different energies, an ongoing development will permit capturing simultaneously up to 8 images at different pre-selected energies for each neutron burst. This will be accomplished by viewing the image formed on the optical amplifier's phosphor by a special intensified CCD camera consisting of an image splitter, and a special segmented, 8-fold independently-gateable image intensifier.

In the FNRR system described above we propose to utilize the fast neutrons produced in the B(d,n) reaction. At a deuteron energy of 3.5 MeV and beam current of 200 μA the yield of the fast neutrons is about $10^{11}$ n/s [10]. We estimate that with several (3-5) 20 cm wide detector arrays, container throughput will be ~10 containers/hr.

### 3.3 Gamma-Resonance Absorption (GRA)

Gamma Resonant Absorption (GRA) is essentially a nitrogen-specific radiographic imaging method, combining highly-penetrating radiation with sensitivity and specificity to nitrogenous explosives. It detects the latter and distinguishes them from benign objects on the basis of nitrogen density, which can be determined with adequate precision from a small number of views [21]. GRA is thus uniquely well-suited to inspecting large, massive objects such as aviation/marine containers, heavy vehicles and rolling stock, the resonant probe being a high energy γ-ray of 9.172 MeV. Among these application scenarios, detecting small threat amounts that are relevant to aviation security may be the one in which the overall characteristics of GRA are best brought into play.

GRA was proposed by Soreq NRC to the U.S. Federal Aviation Administration (FAA) in 1985 [32] and successfully taken under its sponsorship through several experimental feasibility rounds, including a proof-of-principle laboratory test (1989) [33], a blind test on aviation baggage aggregates (1993) [34] and a preliminary run on an aviation container (1998) [35].

These tests were all conducted at existing accelerator facilities, since resonant γ-rays are only produced with adequate yield and spectral quality by 1.746 MeV protons incident on a highly-enriched $^{13}$C target. Since the beam quality varied considerably (being particularly poor in the 1993 blind test), the single factor that contributed most significantly to success was the use of resonant-response detectors (nitrogen-rich, organic liquid scintillators), specially developed at Soreq for this purpose [36] (see below).

The 9.172 MeV level in $^{14}$N was first identified in 1951 [37, 38]. It has spin parity $I^{\pi}=2^+$ and is the lowest T=1 state in this nucleus. The total level width was first measured in 1959 [18], and the present adopted value is:

$$\Gamma_{tot} = 122\pm8\,\text{eV} \quad \leftrightarrow \quad \text{lifetime } \tau = (5.5\pm0.4)\cdot 10^{-18}\,\text{s} \quad [39]$$

The partial γ-decay width is: $\Gamma_{\gamma} = 6.3\pm0.4\,\text{eV}$    of which ~86% goes to the g.s. [40]

The remaining width:    $\Gamma_{p}{\sim}116\,\text{eV}$ represents proton emission (all other channels closed)

Among unbound levels in light nuclei, this level exhibits a very low particle decay width, as evident from the high branching ratio $\Gamma_\gamma/\Gamma_{tot} = 0.052$ (typical values are $10^{-4}$-$10^{-8}$). However, it is this hindered particle decay that produces high GRA nitrogen contrast sensitivity, since the resonant absorption cross-section is also proportional to $\Gamma_\gamma/\Gamma_{tot}$ [41] (Fig. 9).

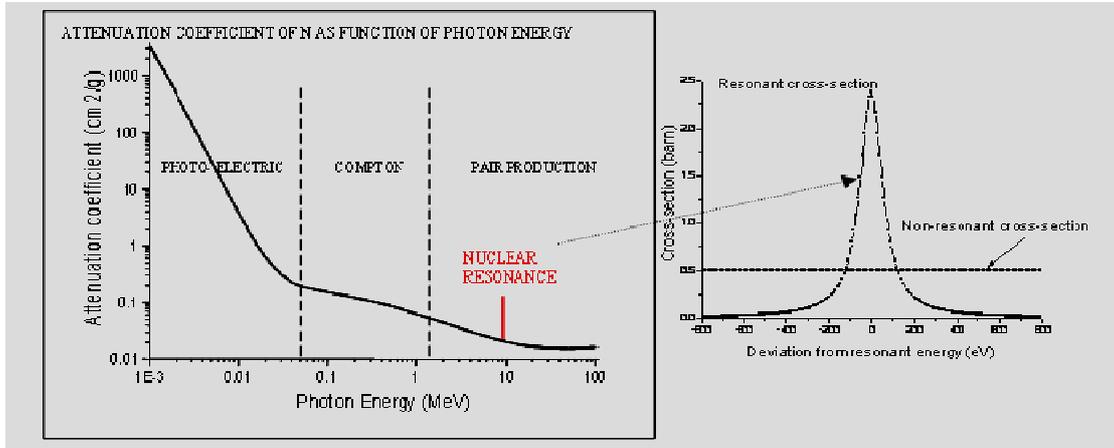

*Fig. 9: Mass attenuation coefficient in nitrogen (left), with horizontal scale expanded to show 9.172 MeV resonance (right).*

For GRA, the optimal 9.172 MeV $\gamma$-ray source is proton capture on $^{13}$C at the 1.746 MeV resonance. For kinematic reasons, resonant flux is confined to a narrow cone around polar angle $\theta_R = 80.7^\circ$ (Fig. 10).

Independently-operating GRA stations can thus be built around its periphery. However, this source is less than ideal, for the following reasons:

**a)** $\gamma$-yield is only ~6·$10^{-9}$/incident-proton [18-20]

**b)** the emission line is broadened from 122 eV (nucl. linewidth) to ~520 eV [19,20]

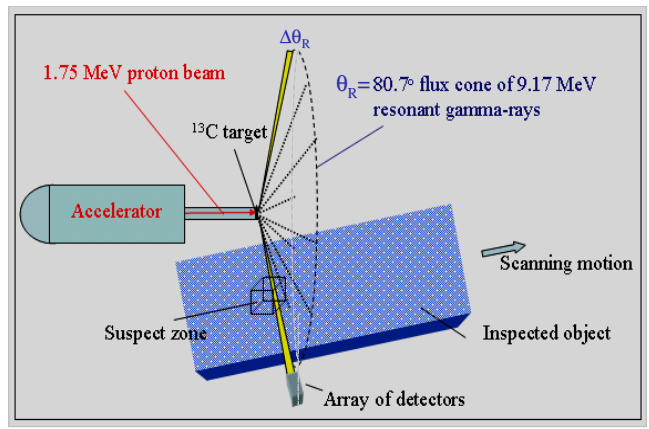

*Fig. 10: Physics-dictated GRA configuration*

One consequence of (b) is that the angular width of the flux cone is broadened to $\Delta\theta_R \sim 0.7^\circ$ (FWHM), which practically limits the achievable GRA spatial resolution.

A more fundamental problem associated with the emission-line broadening is that only a small intensity fraction within the cone (of order 122/520 ~0.23) is on-resonance radiation. Consequently, a conventional-response detector such as NaI, BGO or Ge, measures nitrogen sensitivity that is substantially reduced; in fact, by the inverse of this fraction (a factor of ~4).

In contrast to other groups active in the field, Soreq has circumvented this problem by developing resonant-response detectors, in the form of nitrogen-loaded liquid scintillators. With these, resonant flux is sampled, on an event-by-event basis, by counting photo-protons internally-produced via the resonant absorption reaction $^{14}$N($\gamma$,p), the inverse reaction to $^{13}$C p-capture. This is the same process as occurs in the inspected object, but there, the photo-protons do not emerge – their range in matter is only 10-50 μm. However, the detector always sees the transmission flux attenuation due to nitrogen in the object with undiminished contrast sensitivity, independent of the resonant flux fraction in the γ-ray beam.

A rigorous theoretical analysis based on empirical values has shown that such resonant-response detectors, when loaded with ~20% nitrogen (wt per wt), clearly prevail over their non-resonant counterparts [36]. In fact, the scintillators have been doped with up to ~30% nitrogen.

So far, deployment of a GRA-system has been held up by lack of a compact, robust few-mA ion accelerator. For $H_2^+$ beams of $3\,mA$, a GRA screener is anticipated to detect operationally-relevant amounts of bulk/sheet nitrogenous explosives at throughputs of ~20 containers/hr, with <~2% false-alarms. Due to the inordinately high nitrogen contrast-sensitivity associated with the nuclear resonance, only 100-200 counts per pixel are required in the nitrogen images. Thus, radiation doses to inspected items and environment are more than an order-of-magnitude lower than for any other radiation-based explosives detection system. Moreover, no remanent activity is induced.

## 4. Missing GRA Information

In the brief review of GRA (Section 3.2) it was mentioned that the aperture of the resonant flux cone around polar angle $\theta_R = 80.7^o$ is broadened to $\Delta\theta_R \sim 0.7^o$ − this is presumably due to electron shell excitation processes concomitant with p-capture by the $^{13}C$ nucleus, as has frequently been observed in such low-energy reactions [42].

Prior to the present work, it was not known whether "Coulomb explosion" effects [43] on the $H_2^+$ ion, as it traverses the $^{13}C$ target, would further broaden $\Delta\theta_R$. Indeed, it has been known for many years that such effects appreciably broaden p-capture excitation curves [44].

Clearly, in the case that $\Delta\theta_R (H_2^+) \gtrsim 1.5^o$, as there were grounds to believe [43], GRA spatial resolution would be seriously impaired and a principal advantage of the method lost. It was thus of cardinal importance for the viability of the dual-purpose accelerator and inspection system concept to obtain this missing information (as illustrated on Fig.11)

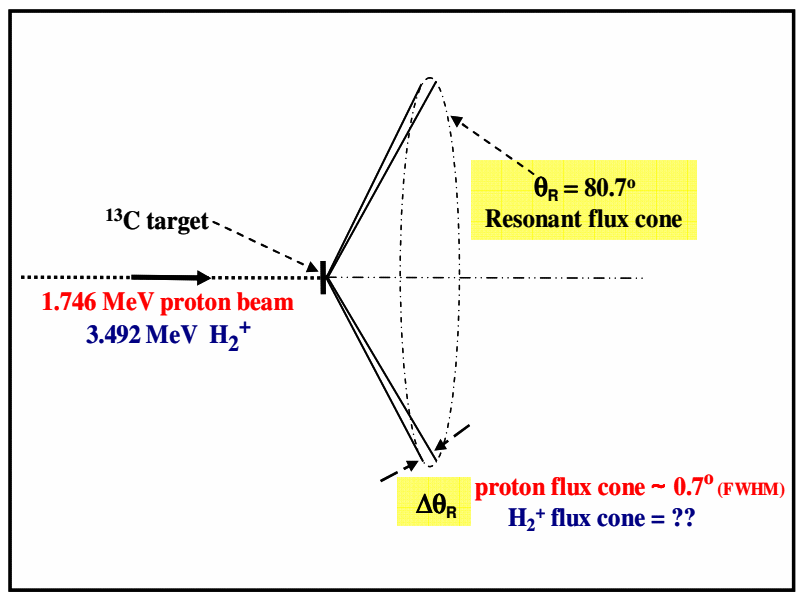

*Fig. 11: Schematic of the missing GRA information for $H_2^+$ beams*

## 5.  GRA Emission-Linewidth Measurements

The experiments were performed at the PTB 3.5 MV single-stage Van-de-Graaff accelerator. With the nitrogen-loaded liquid scintillator (see below), measurements with $H_2^+$ are extremely sensitive to even $10^{-4}$-level deuteron impurities in the beam. In fact, several maintenance rounds (primarily on the gas-manifold that feeds the ion source) and beam tests were required before the experiment could be performed. This issue must also be borne in mind in designing the dedicated mass-2 accelerator, which will need to alternate within tens of seconds between the ionic species and deliver $H_2^+$ beams appropriately low in deuteron contamination.

In resonant absorption studies, geometrical factors should not contribute excessively. This applies to beam optics parameters (divergence, spot-size), and to detector solid angle. In the experiments reported here, beam and detector contributions to the observed 9.17 MeV resonant flux conewidth were $\leq 0.25^o$ and $0.54^o$, respectively. Since the primary interest was whether or not the $H_2^+$ beam would further broaden the conewidth (consistently observed with protons in very tightly-collimated experiments to be $\Delta\theta_R \sim 0.7^o$ [18-20]), geometrical-broadening effects, which add to the intrinsic conewith in quadrature, were indeed sufficiently small.

In practice, the beam was focussed on the 60 µg/cm² thick, self-supporting, highly enriched $^{13}$C target using only the first quadrupole lens on the beamline (located ~6 m upstream), through a pair of adjustable X,Y slits located about halfway from lens to target. Several cm in front of the target, a rectangular collimater limited the beam to apertures of 5 mm and 8 mm in the horizontal (essentially, polar) and vertical (essentially, azimuthal) directions, respectively.

The target was centred on the goniometer axis, on which a resonant-response scintillator [36] (Sect. 3.3) was mounted at a target/detector-face distance of 200 cm. The cocktail contained ~14% nitrogen (wt/wt). Active detector volume was 20 cm long (in the efficiency direction) and 2 cm x 2 cm in cross-section (in the polar and azimuthal directions). It was coupled to a fast 2" Hamamatsu phototube. Photo-protons from the resonant absorption process $^{14}$N$(\gamma,p)^{13}$C were distinguished from non-resonant Compton electron events via pulse shape discrimination.

In the first series of measurements, the 9.17 MeV excitation curves were mapped out with both proton and $H_2^+$ beams. This was performed using an NaI scintillator, by choosing several fixed Van-de-Graaff energies and varying target voltage in small steps over a 16.8 kV range in between them. The procedure is inherently more precise and can be performed in much finer energy increments than the conventional method using the $90^o$ analysing-magnet field calibration. From these curves a relative energy scale was established, the absolute calibration being obtained by associating the known resonance energy with the halfway point of the yield rise at low bombarding energies. From the data on Fig. 12 (Note: energy scale increases from right to left), the working points for the GRA emission linewidth runs were chosen. Fig. 13 shows the angular distribution of photo-protons (normalized to NaI counts) from Run #3.

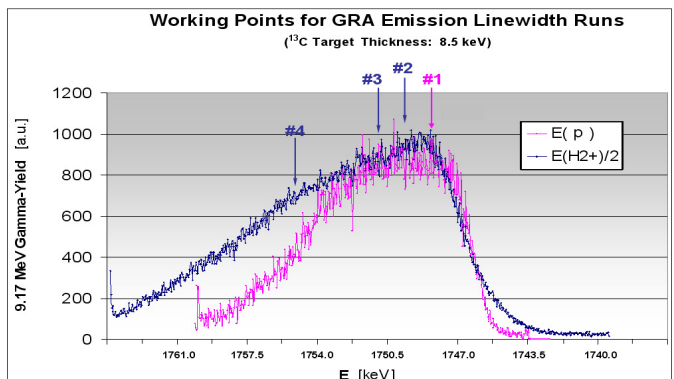

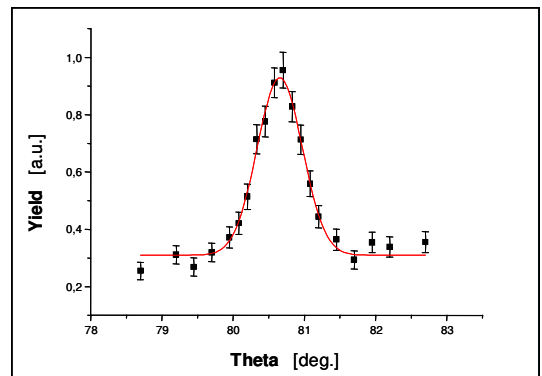

*Fig. 12:  9.172 MeV excitation curves (p & $H_2^+$)*      *Fig. 13: photo-proton distribution (#3)*

It can be seen in Fig. 12 that the $H_2^+$ excitation curve is significantly broader than its proton counterpart, in line with previous observations [44]. The background in Fig. 13 is entirely due to impurities of deuterons in the $H_2^+$ beam, since resonantly-produced photo-protons are indistinguishable from neutron-induced knock-on protons at the same energy (1.51 MeV) in the PSD spectrum. Indeed, in Run #1, taken with protons, the background was non-existent.

In the second series of measurements, photo-proton angular distributions were taken for proton and $H_2^+$ beams at the bombarding energies indicated in Fig. 12. Each distribution was then fitted to a Gaussian with constant background. Fig. 14 shows the fitted FWHM-widths obtained. Evidently, within statistics, no systematic broadening with $H_2^+$ beams is observed, even when the resonance is driven ~8 keV/nucleon into the depth of the target (Run #4).

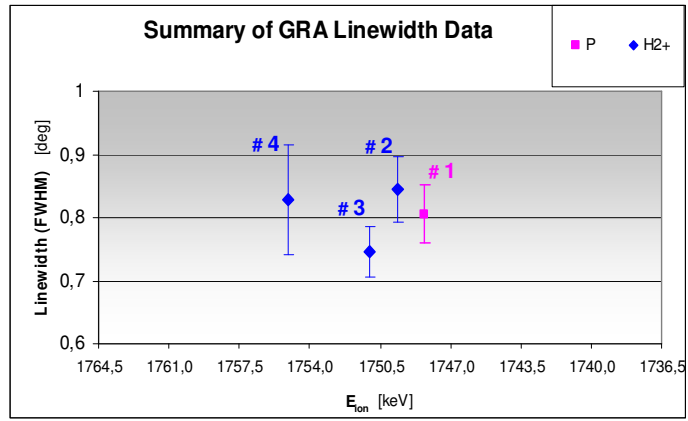

*Fig. 14: Fitted distribution widths (FWHM)*

It is surmised that the absence of an appreciable emission-line broadening effect with $H_2^+$ ions is related to the well-known "wake" effect [43], whereby the trailing particle progressively aligns itself behind the leading one, as they traverse the target layers (prior to one of them being captured by a $^{13}C$ nucleus). Thus, the Coulomb repulsion between the two projectiles has an appreciable longitudinal component, that manifests itself primarily as a broadening of the excitation yield curve, as has been frequently observed for capture resonances populated with $H_2^+$ ions [44]. In contrast, this Coulomb repulsion does not have a large transverse component, and thus does not give rise to appreciable 9.172 MeV emission-line-broadening.

## 7. Summary

The results of the present work have favourable consequences for accelerator beam-energy resolution. In fact, even an energy-spread of up to ~15 keV (FWTM) for the $H_2^+$ beam would be acceptable. Thus, RF-based accelerators (linacs, RFQ's and internal-target cyclotrons), that are more amenable to high-current applications than their electrostatic counterparts, but deliver beams with somewhat poorer energy-resolution, can now be considered as realistic options. Moreover, RF-technologies are generally more easily ruggedized and miniaturized, a crucial advantage for systems required to operate reliably and for extended periods in the field.

The combined cargo inspection system envisaged will automatically and reliably detect small, operationally-relevant amounts of concealed explosives and SNM. It will be cost-effective, employing largely-common hardware, but different reactions and data acquisition modes. The system is also inherently multi-level [2] in the sense that, if an alarm is raised, the cargo item in question is subjected to further screening within the system itself, until it can be definitely cleared or declared suspect. Throughput is projected to be 10-20 aviation containers/hr.

At the heart of the concept is a dual-purpose accelerator for mass-2 ions, alternately delivering several mA (cw) of $H_2^+$ and sub-mA (pulsed) deuterons. As shown by the 9.172 MeV emission line data of the present work, beam quality requirements are only moderately stringent, which opens up the field to accelerator technologies that are compact and robust enough to function for extensive periods under harsh environmental conditions. Thus, the present concept outlines a realistic accelerator and associated security-screening system that respond to contemporary requirements in the cargo-inspection scenario.